\documentstyle[12pt]{article}

\begin{document}
\baselineskip 18pt

\begin{center}
{\Large\bf Low temperature nonequilibrium dynamics

in transverse Ising spin glass}\\

\bigskip

{G. Busiello$^a$\footnote{Corresponding
author,tel.(0)39-89-965426 -
fax (0)39-89-965275 - e-mail: busiello@sa.infn.it}}\\

\medskip

{\small\it Dipartimento di  Fisica ``E.R. Caianiello",
Universit\`a
di Salerno,\\
84081 Baronissi - Salerno and INFM - Unit\`a  di Salerno, Salerno, Italy}\\
\medskip
{R. V. Saburova and V. G. Sushkova}\\
{\small\it Kazan State Power-Engineering University, Kazan, Russia}
\end{center}

\bigskip
{\begin{abstract} The real part of the time-dependent $ac$
susceptibility of the short-range Ising spin glass in a transverse
field has been investigated at very low temperatures. We have used
the quantum linear response theory and domain coarsening ideas of
quantum droplet scaling theory. It is found that after a
temperature quench to a temperature which is less than the spin
glass transition temperature the $ac$ susceptibility decreases
with time elapsed after the initial quench towards equilibrium
approximately logarithmically. It is shown that the transverse
field of "tunneling" has nonessential effect to a nonequilibrium
dynamical properties of the magnetic droplet system. It is found
that the time dependence of $ac$ susceptibility has a qualitative
agreement with some experimental results.\end{abstract}}

\bigskip \bigskip \bigskip \bigskip \bigskip


\newpage

\section{Introduction}

Aging phenomena and nonequilibrium slow dynamics have been
investigated during last years in many materials with glassy
properties such as spin glasses, polimer glasses, orientational
glasses, simple liquids like glycerol and gels [1-12]. These
systems are characterized by the existence of a nonequilibrium low
temperature phase and aging [13-25]. Despite a great progress
towards the understanding of nonequilibrium dynamics, some
problems remain open. One of them is an investigation of a very
low temperature nonequilibrium dynamics in quantum spin glasses,
namely the nature of quantum channels of relaxation, the behavior
of quantum glassy system subjected to periodic driving force,
aging at very low temperatures. The natural basis for the
interpretation of aging is based on coarsening ideas of a slow
domain growth of a spin-glass type ordered phase [6, 9, 15]. For
theoretical studies of quantum fluctuations in disordered media
there is a variety of techniques including replica theory,
renormalization group, Monte Carlo simulations, the Schwinger and
Keldysh closed-time path-integral formalism and others [26-38]. A
large attention in the last decade was payed to the spin glasses
representing a model systems for study of nonequilibrium dynamics
[39-47].

In this paper we investigate the real time nonequilibrium dynamics
in $d$ - dimensional Ising spin glass in a transverse field in
terms of droplet model at very low temperatures. We calculate the
$ac$ susceptibility as a function of the time elapsed since a
thermal quench. We show that quantum effects alter the
nonequilibrium dynamics in the spin glass phase at very low
temperatures.

In classical spin glasses in the $ac$ susceptibility measurements
the magnetic response of the system to small $ac$ magnetic field
after quenching shows aging effects. This response depends on its
thermal history and the time interval the system has been kept at
a constant temperature in the glass phase.

Spin glass materials have nonequilibrium dynamics and provide a
measure of processes causing the aging. It is assumed that
isothermal aging is a coarsening process of domain walls, and the
temporal $ac$ susceptibility (real part $\chi'$ and imaginary part
$\chi''$) at a given frequency of $ac$ magnetic field $\omega$ at
time $t$ after the quenching scales as [17, 24, 35]

\begin{equation} \frac{\chi '' (\omega,t)-\chi_{eq} ''(\omega)}{\chi ''
(\omega,t)}\sim \left [\frac{L\left(1/
\omega\right)}{L(t)}\right]^{d-\theta},\end{equation}
\begin{equation} \frac{\chi ' (\omega,t)-\chi_{eq} '(\omega)}{\chi '
(\omega,t)}\sim \left [\frac{L\left(1/
\omega\right)}{L(t)}\right]^{d-\theta}\end{equation} for $| {\rm
ln}\, \omega|\ll {\rm ln}\, t$ if $L(t)$ is proportional logarithm
of $t$; $\theta \leq (d-1)/2$. Here $L\left(1/ \omega\right)$ is
the typical size of the droplet being polarized by the $ac$ field,
$L(t)$ is the typical domain size. $L$ may change according to
logarithmic growth law $(L \sim {\rm ln}\, t)$ or algebraic law
$(L\sim t^\alpha)$ [39, 45]. The relation $\left [L\left(1/
\omega\right)/L(t)\right]$ is proportional to $\left[{\rm ln}\,
(\omega/\omega_0)/{\rm ln}\, (\tau/\tau_0)\right]^{1/\psi}$ if the
droplet picture is used, $\psi$ is a some exponent, $0\leq\psi\leq
d-1$, $\omega_0=\tau_0^{-1}$, $\tau_0$ is a certain microscopic
characteristic unit of time. The logarithmic growth law (like
algebraic growth law) is supported by recent experiments [17, 45].
The expressions (1) - (2) were found when the relaxation is
governed by thermal activation over a free-energy barrier $B$. The
barrier for annihilation and creation of the droplet excitations
are assumed to scale as $B\sim \Delta L^\psi$; $\Delta$ is a
barrier energy at $T\ll T_g$. The barriers have a broad energy
distribution. A droplet with $B$ lasts for a time $\tau$ of order
$\tau\sim\tau_0 {\rm exp} \left[B/(k_B T)\right]$ where $k_B$ is
the Boltzmann constant. $\tau$ is the rate of classical activation
over energy barrier $B$.

After time $t$ after quenching the typical linear domain size of
the system has accordingly grown to $R(t)\sim\left[(k_B
T/\Delta(T)) {\rm ln} \left(t/\tau_0\right) \right]^{1/\psi}$. In
the $ac$ susceptibility measurements at angular frequency
$\omega$, the $ac$ field excites droplets of length scales up to
$L\left(1/\omega\right)\sim\left[(k_B T/\Delta(T))\, |{\rm ln}
\left(\omega/\omega_0\right)| \right]^{1/\psi}$. Because elapsed
time $t\gg \omega^{-1}$ we have $L\left(1/\omega\right)<R(t)$.
These droplets have walls which partly coincide with walls of the
domain of size $R$. The presence of such frozen-in domain wall
influences the small scale droplet excitations. Some droplets
which touch it can reduce their excitation gap, compared with
others in the bulk of domains. In the presence of domain wall at a
typical distance $R$ from each other D.~Fisher and D.~Huse have
found the free energy gap of a droplet of size $L$ in the
following form [35]

\begin{equation} F_{L,R}^{typ}=\gamma_{eff}\left[\frac{L}{R}\right ]
\left(\frac{L}{L_0}\right)^\theta,\,\,\,\, L<R.\end{equation} with
effective stiffness constant $\gamma_{eff}\left[L/R\right ]=\gamma
\left[1-c_\nu \left(L/R\right )^{d-\theta}\right]$, $\gamma_{eff}$
is a function of the ratio $(L/R)$. $L_0$ is a certain microscopic
unit of length, a short-distance cutoff; $c_\nu$ is a constant
independent of time and frequency which happens to be anomalously
small. So within the droplet picture, aging proceeds by coarsening
of domain walls as usual phase ordering processes. The domain wall
serves as the frozen-in extended defect for the droplets and
reduces their stiffness constant. It is known that the
susceptibility is inversely proportional to $\gamma$. It was
derived, for example, for the real part of $\chi (\omega,t)$ that
[24]

\begin{equation} \chi '(\omega,t)=\chi '_{eq}(\omega)
\left(1-c_\nu \frac{\Delta
\gamma}{\gamma}\right)^{-1}\end{equation} where $\Delta
\gamma/\gamma\sim \left(L(1/\omega)/R(t)\right)^{d-\theta}$ [35].
$\chi '_{eq}(\omega)$ is the real part of the equilibrium
susceptibility.

All aforementioned dynamical processes (droplet excitations,
thermal activation time and so on) are thermally activated
processes. We shall call these processes as classical processes in
classical regime when $\beta \Gamma_L \ll 1$. In quantum regime
$\beta \Gamma_L \gg 1$ [36, 37], $\beta=(k_B T)^{-1}$. For quantum
droplet model developed in [36-38] $L(1/\omega)\sim
\left[(1/\sigma) \left|{\rm
ln}\left(\Gamma_0/\omega\right)\right|\right]^{1/d}$ and $L(t)\sim
\left[(1/\sigma) \left|{\rm ln}\left(\Gamma_0 t
\right)\right|\right]^{1/d}$, where the coefficient $\sigma$
varies a little from droplet to droplet. So we suppose the
logarithmic growth law of the time dependent length scale $L(t)$
of droplet excitations within the frozen-in defect at scale $R$.
If we take into account the correction to the true equilibrium
which occurs due to the finite domain size $R$ we have instead
$\gamma$ to write $\gamma_{eff}$. Instead of thermal relaxation
time $\tau$ for classical process we use in our quantum case the
quantum tunneling rate $\Gamma_L$ for a droplet of linear size
$L$. So the length scale $L(t)$ growth is determined not by the
thermal activation over the free energy barriers between minima
but by quantum fluctuations which cause a droplet tunneling
through the barrier at rates that do not vanish for $T\rightarrow
0$. Thus the quantum fluctuations become dominant at very low
temperatures. The fraction of droplets which are
quantum-mechanically active at $T\rightarrow 0$ is proportional to
$\Gamma_L$.

In general, quantum effects could change nonequilibrium dynamics
in glassy phase at very low temperatures. The understanding aging
dynamics in quantum system requires a quantum-dynamical approach.

The paper is organized as follows. In next section we give the
definitions and main properties of spin glass droplet model. In
section 3 we give the linear response formalism and general
expression for $ac$ magnetic susceptibility. In section 4 we
present a summary of our results and conclusions.

\section{The model and Hamiltonian}

In this paper we use a phenomenological quantum droplet model of
spin glass theory [35-37] (which does not use the mean-field
approximation) in order to describe the nonequilibrium behavior of
the magnetic dynamical susceptibility at very low (but finite)
temperatures $T$ $(T\leq 1K)$. We use the quantum Hamitonian for
the short-range Ising spin glass in a transverse field. This model
Hamiltonian may be  appropriate for experimental systems such as
dipolar magnet ${\rm LiHo}_x{\rm Y}_{1-x}{\rm F}_4$, proton
glasses, alcali halides with tunneling impurities and other
quantum systems [9, 36-38].

We shall be interested in the very low temperatures in the ordered
spin glass phase and ignore critical effects. The transverse
field, applied perpendicular to the Ising axis, introduces
additional quantum channels of relaxation at that, for example,
depressing the spin-ordering temperature $T_g$.

The droplet model describing the low-dimensional short-range Ising
spin glass is based on renormalization group arguments [35, 36].
In dimensions above the lower critical dimension $d_l$ (usually in
spin glass $2\le d_l <3$) the droplet model finds a low
temperature spin-glass phase in zero magnetic field. This phase
differs essentially from the spin-glass phase in the mean-field
approximation of the Sherrington-Kirkpatrick infinite-range
spin-glass model [9]. In the droplet model there are only two pure
thermodynamical states related to each other by a global spin
flip. In magnetic field there is no phase transition. A droplet is
an excited compact cluster in an ordered state where all the spins
are inverted. The natural scaling ansatz for droplet free energy
$\epsilon_L$ (which are considered to be independent random
variables) is $\epsilon_L \sim L^\theta, L\ge \zeta (T)$; $\zeta$
is the correlation length, $L$ is the length scale of droplet and
$\theta$ is the zero temperature thermal exponent. One droplet
consists of order $L^d$ spins. Below $d_l$, $\theta<0$; above
$d_l$ one has $\theta>0$. The droplet excitations have a broad
distribution of their free energies with the probability
distribution  $P_L (\epsilon_L) d\epsilon_L$   of droplet free
energies at scale $L$ for large $L$ in a scaling form [4, 6]
\begin{equation}
P_L (\epsilon_L) d\epsilon_L = {d\epsilon_L \over \gamma (T)
L^\theta} {\cal P} \left( \epsilon_L \over \gamma (T)
L^\theta\right),\, L\to\infty\ .
\end{equation}
It is assumed that $P_L (x\to 0) >0$, ${\cal P}_L(0)-{\cal
P}_L(x)\sim x^\phi$ at $x\to 0$; $\phi\leq1$. $\gamma(T)$ is the
stiffness constant of domain wall on the boundary of the droplet
which is of order of characteristic exchange ${\cal I}=
\overline{\left({\cal I}_{ij}^2\right)}^{1\over 2} $ at $ T=0$ and
vanishes for $T\ge T_g$.

The Hamiltonian of the $d$ - dimensional quantum Ising spin glass
in a transverse field is given by
\begin{equation}
{\cal H} = -\sum_{i,j} {\cal I} _{ij} S^z_i S^z_j - \Gamma \sum_i
S^x_i
\end{equation}
where $S_i$ are the Pauli matrices for a spin at site $i$.
$\Gamma$ is the strength of the transverse field. The sum in (6)
is performed over nearest neighbors. The interactions ${\cal
I}_{ij}$ are independent random variables of mean zero and
variance ${\cal I}= \overline{\left({\cal
I}_{ij}^2\right)}^{1\over 2} $. This model shows, for example, the
physics of proton glasses, mixed betaine phosphate-phosphite [9].
The transverse field may be interpreted as the frequency of proton
tunneling. The quantum spin glass transition in a dilute dipole
coupled magnet ${\rm LiHo}_x{\rm Y}_{1-x}{\rm F}_4$ [9] also is
described by this model Hamiltonian. The properties of model (6)
in mean-field approximation have been studied in many papers [6,
9]. It was found that there is a high-temperature critical
behavior at the temperature $T_c(\Gamma)\sim{\cal
I},\,\Gamma\ll{\cal I}$ and low-temperature critical behavior with
the zero-temperature critical point $T_c({\cal
I})=0,\,\Gamma_c(0)={\cal I}$, where $\Gamma_c$ is the critical
value of $\Gamma$ below which the spin-glass phase can exist. We
suppose as in our early papers [37, 38] that quantum system with
the Hamiltonian (6) has a true glass phase transition.

We shall use the quantum model Hamiltonian (6) of the $d$ -
dimensional Ising spin glass in the transverse field. If
$\Gamma=0$ the Hamiltonian (6) is the $d$ - dimensional classical
Ising spin glass. One can use the Suzuki - Trotter formalism [36]
to show that the $d$ - dimensional quantum mechanical system (6)
is equivalent to a classical statistical mechanical system in
$(d+1)$ - dimensions (the extra dimension is imaginary time) with
the classical Hamiltonian

\begin{equation}{\cal H}_{cl}=\Delta\tau \sum_{k=1}^{L_\tau}
\sum_{\langle ij\rangle} {\cal I}_{ij} S_{i,j} S_{j,k}-{\cal I}_F
\sum_{k=1}^{L_\tau} \sum_i S_{i,k} S_{i,k+1}\end{equation} where
the $S_{i,k}=\pm 1$ are the classical Ising spins, representing
the $z$-component of the quantum spins at site $i$ and imaginary
time $\tau=k\Delta \tau$, the indices $i$ and $j$ running over
sites of the $d$ - dimensional lattice. The imaginary time
direction has been divided into $L_\tau$ time slices of width
$\Delta\tau$. The calculation gives that ${\rm exp}\left({\cal
I}_F \right)={\rm tanh}\left(\Delta\tau\Gamma\right)$, $\Delta\tau
L_\tau=(k_B T)^{-1}$, here and throughout the paper we use units
where $\hbar=1$. Using properties of the Hamiltonian (7) of the
$(d+1)$ - dimensional classical model at temperature $T=0$
M.~J.~Thill and D.~A.~Huse [36] have assumed that in each of the
two ordered states (for $\Gamma<\Gamma_c$ and $T<T_g$) for enough
low $T$ and appropriate range of length scales $L$ the quantum
Hamiltonian (6) can be represented as independent quantum
two-level systems (low energy droplets) with the Hamiltonian

\begin{equation}
{\cal H} ={1 \over 2} \sum^{\sim}_L \sum_{D_L} \left( \epsilon
_{D_L} S^z_{D_L} +\Gamma_L S^x_{D_L} \right)
\end{equation}
where $S^z_{D_L} $ and $S^x_{D_L}$ are the Pauli matrices
representing the two states of the droplet; the sum is over all
droplets $D_L$ at length scale $L$ and over all length scales
$L$, and
\begin{equation} \sum^{\sim}_L\sim \int^\infty_{L_0}
{dL\over L}
\end{equation}
where $L_0$ is a short-distance cutoff. $\epsilon_{D_L}$ is the
droplet energy which is independent random variable. The droplet
length scale $L$ is more or of order of the correlation length
$\zeta$.
\begin{equation}
\Gamma_L=\Gamma_0 \exp[-\sigma L^d]
\end{equation}
is the tunneling rate for a droplet of linear size $L,\,\Gamma_0$
is the microscopic tunneling rate; $\sigma$ is the surface tension
for the interface between the two droplet states, $\sigma$ is
approximately the same for all droplet. We will assume $\Gamma_L$
is the same for all droplets of scale $L$ [36]. The Hamiltonian of
a single droplet is the $2\times 2$ matrix
\begin{equation}
{1\over 2} \left(
\begin{array}{cc}
\epsilon_{D_L} & \Gamma_L \\
\Gamma_L & -\epsilon_{D_L}
\end{array} \right)
\end{equation}
with eigenvalues $E_\pm =\pm \sqrt{\epsilon^2_L + \Gamma^2_L}$.
$E=2|E_\pm|$ is the energy difference between the two eigenvalues.
Note the Hamiltonian (9) is similar to the Hamiltonian of
two-level system in real glass [32].

In the quantum droplet model of M.~J.~Thill and D.~A.~Huse [1] the
relative reduction of the Edwards-Anderson order parameter
$q_{EA}(T)$ from its zero-$T$ value $q_{EA}(0)$ for $\theta>0$ and
$L^*(T)\geq\zeta$ is given by
\begin{equation}1-\frac{q_{EA}(T)}{q_{EA}(T=0)}
\sim\frac{k_B T}{\gamma L^{*\theta}(T)}=\frac{k_B
T}{\gamma}\frac{\sigma^{\theta/d}}{\ln^{\theta/d}(\Gamma_0 /(k_B
T))}\,\,\,\,\,\,\,\, (T\rightarrow 0)
\end{equation}
where the crossover length scale is

\begin{equation}L^*(T)=\left({1\over \sigma} \ln
{\Gamma_0 \over k_B T} \right)^{1\over d}.\end{equation}

For droplets with $L\ll L^*(T)$ and $\Gamma_L \gg k_BT$ the energy
$\sqrt{\epsilon^2_L+\Gamma^2_L}$ is always greater than $k_BT$ and
thermal fluctuations are therefore not essential at temperature
$T$. These droplets behave quantum-mechanically. The larger
droplets ($L\gg L^*(T)$) have $\Gamma_L \ll k_BT$ and behave
classically. Those large droplets ($\epsilon_L\le k_B T,\,
\Gamma_L \le k_BT$) are thermally active and reduce $q_{EA}$. For
$\theta>0$ this reduction is dominated by smallest droplets. Below
$d_l$ $(\theta<0)$ the thermally-excited droplets disorder the
system reducing $q_{EA}$ to zero at any temperature.

There is complicated dynamical-to-quantum crossover depending on
temperature, frequency of $ac$ field and length scale $L$.
According to [36] the crossover dynamical length is determined
from condition $\Gamma_L^{-1}=t$, i. e.

\begin{equation}L_{dyn}^*(T)\sim \left(\frac{\sigma}{\Delta}k_B
T\right)^{\frac{1}{\psi-d}}.\end{equation} The system behaves
presumably classically or quantum-mechanically when the dominant
length scale $L$ is above or below $L_{dyn}^*$ for fixed frequency
$\omega$. When the droplets behave quantum-mechanically they have
a characteristic rate - the Rabi frequency (is of order
$\Gamma_L$).

In this paper we neglect droplet-droplet interaction and
droplet-lattice one.

\section{Linear response}
We consider the time-dependent Hamiltonian $\hat{\cal H}$ of the
quantum system in the form [48]

\begin{equation}
\hat{\cal H}=\hat{\cal H}_{0}+\hat{\cal H}'(t)=\hat{\cal
H}_{0}-\hat A h(t).
\end{equation}
where $\hat{\cal H}_{0}$ is nonperturbated Hamiltonian of system,
$\hat{\cal H}_{0}\neq\hat{\cal H}(t)$ and suppose that the
external perturbation $\hat{\cal H}'(t)$ is in some sense small.
$\hat{\cal H}_0$ describes the equilibrium system. Here the $\hat
A$ is the linear operator through which the external time varying
force $h(t)$ couples to the system.

We use the quantum-mechanical calculation for the system dynamical
response $\Delta \hat B (t)\equiv
\langle\hat{B}(t)\rangle-\langle\hat{B}\rangle_{0}$ to the force
$h(t)$ in terms of the time-evolution operator $U(t,t')$; $\hat
B(t)$ is the Heisenberg operator, $\hat B(t)=\hat U^{\dag}(t,t')
\hat B(t_0) \hat U(t,t')$, $\langle \hat B\rangle_0$ is the
equilibrium expectation value of $\hat B$. It is necessary to make
approximation for $\hat U(t,t')$ carrying out the well-known
perturbation expansion of $\hat U(t,t')$ through first order in
applied force in the following form

\begin{equation}
\hat{U}(t,t') \simeq \hat{U}_{0}(t,t') \left\{ \hat 1-{i\over
\hbar}\int^{t}_{t'}dt_{1} \hat{U}^{\dagger}_{0}(t_{1},t')\hat{\cal
H}' (t_1)\hat{U}_{0}(t_{1}, t') \right\}
\end{equation}
where $\hat U_0(t,t')=\exp \left [-\frac{i}{\hbar} (t-t') \hat
{\cal H}_0\right]$, the sign $\dag$ means the conjugate value.

We take into account the first- and second-order linear response
functions for the linear response functional of the form [48]

\begin{equation}\Phi(t,t';t_0)\equiv {1 \over i\hbar}\langle[\hat{A}(t_0),
\hat{B}(t,t')]\rangle_{0}
\end{equation} where $\langle \ldots\rangle_0$ means the thermal
average with the density matrix $\hat{\rho}_0=\hat{\rho}(t_0)$,
where $t_0$ is the time moment when the perturbating field is
turned on, $\hat{\rho}_0=({\rm Tr}\, \exp[-\beta \hat {\cal
H}])^{-1}$ $\times \exp[-\beta \hat {\cal H}]$.

Now we apply the aforementioned dynamical response relations to
our magnetic droplet system. Then the response
$\langle\hat{B}(t)\rangle$ represents the induced magnetization of
droplet system $M(t)$. $\langle\hat{B}\rangle_{0}$ is the
equilibrium magnetization $M_0$. If in $z$-direction a small
magnetic oscillating field $h(t)=h\exp[i \omega t]$ is applied,
the induced magnetization (in $z$-direction) of the sample is

\begin{equation}M(t)=[\chi'(\omega)\cos(\omega t)+\chi''(\omega)
\sin(\omega t)]hV,
\end{equation}
where $\chi(\omega)=\chi'(\omega)-i\chi''(\omega)$ and $V$ is the
sample volume, $\chi_{zz}=\chi$; $h$ and $\omega$ is the amplitude
and the angular frequency of $ac$ field.

In order to observe a history dependence and aging in spin glass
the sample is quenched infinitely fast at zero $dc$ magnetic field
from temperature $T\gg T_g$ to the temperature $T_1<T_g$ which is
reached at time $t=0$. At this moment a very small external
magnetic oscillating field $h(t)$ is applied to measure the $ac$
susceptibility of the sample. The evolution continues in
isothermal conditions, $\chi_{ac}$ is measured as a function of
the time $t$ elapsed since the sample reached the temperature
$T_1$ at fixed frequency $\omega$.

If the field changes are small the magnetization $M(t)$ responses
linearly to a magnetic field $h(t)$ applied at once when the
quench was finished, i. e. at time moment $t=0$. The system is
probed at the time $t$ after quench end (the "age"). Using linear
response theory the magnetization of droplet system is [25]

\begin{equation}M(t)-M_0=\int_0^t \chi(t,t_1) h(t_1)\, dt_1=\int_0^t \chi(t,t-t') h(t-t')\,
dt' \end{equation} where $\chi(t,t-t')$ is the magnetic dynamical
susceptibility (the dynamical response functional including the
first- and second-order linear response functions). $\chi(t,t-t')$
determines the magnetic response at time $t$ to a unit magnetic
field impulse at time $(t-t')$. The nonequilibrium processes are
probed by the low-frequency $ac$ susceptibility measurements. The
frequency dependent $ac$ susceptibility is measured by applied
$ac$ magnetic field $h(t)$ at time $t=0$. Then $\chi(\omega,t)$
may be found by the Fourier transform of the magnetization over
time segment $t_m$ ($t_m\sim 2 \pi/\omega$) centered around $t$
[25, 39]

\begin{equation}
\chi(\omega,t)={1 \over t_m} \int_{t-{t_m \over 2}}^{t+{t_m \over
2}}dt'' e^{-i \omega t''} \int_0^{t''}dt' \chi(t'',t''-t') e^{i
\omega (t''-t')}
\end{equation}
If the magnetic response function varies little over time segment
$t_m$ the susceptibility $\chi(\omega,t)$ will be equal to [25]

\begin{equation}\chi(\omega,t)=\int_0^t dt' \chi(t,t-t')e^{-i \omega t'}
\end{equation}
The in-phase component of the $ac$ susceptibility
$\chi'(\omega,t)$ can be written

\begin{equation}
\chi'(\omega,t)={\rm Re}\left[\int_0^t dt' \chi(t,t-t')e^{-i
\omega t'}\right].\end{equation} The out-of-phase component of the
$ac$ susceptibility is given by

\begin{equation}\chi''(\omega,t)={\rm Im}\left[\int_0^t dt' \chi(t,t-t')e^{-i \omega
t'}\right].\end{equation}

We consider the behavior of the magnetic droplet system described
by the Hamiltonian $\hat {\cal H}$ under $ac$ field $h(t)$ in
quantum regime $(\Gamma_L\gg k_B T)$ when the droplet excitation
energy $\sqrt{\epsilon_L^2+\Gamma_L^2}$ is always greater than
$k_B T$, i. e. the temperature is too small to thermally excite
the droplets which behave classically and thermally active
(classical regime). In our calculations below we assume that
$\theta>0$ $(d>d_l)$.

Following to aforementioned quantum droplet theory and domain
growth ideas we have the following expression for susceptibility
$\chi_{D_L}$ ($\chi_{D_L}$ is proportional to the contribution of
a single droplet to the $ac$ susceptibility) up to some factor
$\sim q_{EA}^2 L^{2 d}$

$$ \chi_{D_L}(\omega,t)-\chi_{D_L}(\omega=0)\sim -\tanh
\left(\beta a_L/2\right) \sin ^2 \varphi
\frac{a_L}{a_L^2-\omega^2} $$ $$ -h\tanh \left(\beta a_L/2\right)
\cos\varphi \sin ^2 \varphi \left(\left\{\frac{(7 a_L^2-10
\omega^2)\cos(\omega
t)}{(a_L^2-\omega^2)(a_L^2-4\omega^2)}\right.\right.$$
$$\left.-\frac{3 a_L\sin(\omega t) \sin(a_L t)+6 \omega
\cos(\omega t) \cos(a_L t)}{\omega
(a_L^2-4\omega^2)}-\frac{\cos(a_L t)}{a_L^2-\omega^2}\right\}$$
$$+i \left\{\frac{3(a_L^2+2 \omega^2)\sin(\omega
t)}{(a_L^2-\omega^2)(a_L^2-4\omega^2)}\right.$$
\begin{equation}\left.\left.-\frac{3
a_L\cos(\omega t) \sin(a_L t)-6 \omega \sin(\omega t) \cos(a_L
t)}{\omega (a_L^2-4\omega^2)}+\frac{3 \sin(a_L t)}{\omega (
a_L^2-\omega^2)}\right\}\right),
\end{equation}
where $a_L=\sqrt{\epsilon_L^2+\Gamma_L^2}$,
$\sin\varphi=\Gamma_L/a_L$, $\cos \varphi=\epsilon_L/a_L$,
$\chi_{D_L}(\omega=0)$ is the static susceptibility of droplet
$D_L$. The expression (24) was obtained with condition $\omega
t\geq 1$ (and $\Gamma_L<\omega$) because this condition is used
for observing nonstationary dynamics in $\chi_{ac}$ measurements
[12]. If $\omega t\geq 1$, i. e. for low frequencies, the aging
part usually strongly contributes. For high frequencies, $\omega
t\rightarrow \infty$, the aging term does not contribute. Now we
have to average the susceptibility (24) over droplet energies
$\epsilon_L$ and over droplet length scales $L$. In order to
average over droplet energies we use the distribution of free
energies $P_L (\epsilon_L)$ (5). In this distribution we assume
$\phi=0$. We approximate in (24)
$\tanh(\sqrt{\epsilon_L^2+\Gamma_L^2}/2)\simeq 1$ and integrate
like in papers [36, 37, 38]. As the result of the droplet energy
averaging we received the averaged contribution of all droplets of
the system to the real part of susceptibility in the following
form

$$\chi'_L(\omega,t)-\chi'_L(\omega=0)\sim\frac{\Gamma_L^2}{\gamma
L^\theta \omega \sqrt{\omega^2-\Gamma_L^2}} \left( \ln
\frac{\omega+\sqrt{\omega^2-\Gamma_L^2}}{\omega-\sqrt{\omega^2-\Gamma_L^2}}-
\ln \frac{\omega p_L+\sqrt{\omega^2-\Gamma_L^2} b_L}{\omega
p_L-\sqrt{\omega^2-\Gamma_L^2} b_L}\right) $$

$$+\frac{\Gamma_L^2 h}{\gamma L^\theta \omega^3}\left[\sin(\omega
t) \left({3 \over 2} {\rm si}\,(b_L t)+{\rm si}(t (b_L-\omega))+
{\rm si}(t (b_L+\omega))+{3 \over 4} {\rm si}\,(t
(b_L-2\omega))\right. \right.$$

$$\left.+{3 \over 4} {\rm si}\,(t (b_L+2\omega))
\right)+\cos(\omega t) \left(-{3 \over 4} {\rm ci}\,(t
(b_L-2\omega))+{3 \over 4} {\rm ci}\,(t
(b_L+2\omega))-\ln\frac{b_L+\omega}{b_L-\omega}\right.$$

$$\left.-{3 \over 4} \ln\frac{b_L+2\omega}{b_L-2\omega}- {\rm
ci}\,(t (b_L-\omega))+{\rm ci}\,(t (b_L+\omega))+{5\omega \over
b_L}-{3\omega \over b_L}\cos(b_L t)-3 \omega t {\rm si}\,(b_L
t)\right)$$

\begin{equation}\left.-{2\omega \over b_L}\cos(b_L t)-2 \omega t {\rm si}\,(b_L
t)\right]\end{equation} where $\chi'_L(\omega=0)$ is the static
susceptibility of the system of droplets of size $L$,
$b_L=\sqrt{p_L^2+\Gamma_L^2}$, $p_L=2 \omega+\Gamma_L$, ${\rm
si}(\alpha)$ is the sine integral and ${\rm ci}(\alpha)$ is the
cosine integral. Further we average expression (26) over length
scales $L$. While integrating over $L$ we see that the real part
of the susceptibility is dominated by droplets of length scale
$L_1\sim[(1/\sigma)\ln(\Gamma_0 / \omega)]^{1/d}$. $L_1$ is the
natural length scale of the problem when $\Gamma_L\sim\omega$ and
the lower limit of integration over $L$. The upper limit of
integration we took as $L_2\sim[(1/\sigma)\ln(t_0\Gamma_0
)]^{1/d}$. Hence we obtain after averaging over $L$ the following
expression for the real part of the $ac$ susceptibility of droplet
system

$$\chi'(\omega,t)-\chi'(\omega=0) \sim \frac{\Gamma_0^2}{\gamma
\omega^2} \left( (\ln {4\omega^2 \over \Gamma_0^2} -2 )
(2\sigma)^{\theta \over d} d^{-1} {\rm G}\left[-{\theta \over
d}\,,2 \left|\ln{\Gamma_0 \over \omega}\right|\right]\right.$$

$$+{\Gamma_0^2 \over 4\omega^2} (2\ln {4\omega^2 \over \Gamma_0^2}
-1)(4\sigma)^{\theta \over d} d^{-1} {\rm G}\left[-{\theta \over
d}\,,4 \left|\ln{\Gamma_0 \over \omega}\right|\right]+{3\Gamma_0^4
\over 8\omega^4}(6\sigma)^{\theta \over d} d^{-1} {\rm
G}\left[-{\theta \over d}\,,6 \left|\ln{\Gamma_0 \over
\omega}\right|\right]$$

$$\left.+\sigma\left(2 (2\sigma)^{{\theta \over d}-1} d^{-1} {\rm
G}\left[1-{\theta \over d}\,,2 \left|\ln{\Gamma_0 \over
\omega}\right|\right]+{\Gamma_0^2 \over
\omega^2}(4\sigma)^{{\theta \over d}-1} d^{-1} {\rm
G}\left[1-{\theta \over d}\,,4 \left|\ln{\Gamma_0 \over
\omega}\right|\right]\right)\right)$$

$$+{\Gamma_0^2h \over \gamma\omega^3}\left( \left({1 \over \omega
t} ({3 \over 16}\sin(3 \omega t)-{17 \over 12}\sin(2\omega t)-{3
\over 4}\sin(\omega t)\cos(2 \omega t))+\cos(\omega
t)\right.\right.$$

$$\left.\times({5 \over 2}-\ln 3-{3 \over 4}\ln{4\omega \over
\Gamma_0})\right)\left(-d^{-1}(2\sigma)^{\theta \over d}({\rm
G}\left[-{\theta \over d}\,,2 \left|\ln
(t\Gamma_0)\right|\right]-{\rm G}\left[-{\theta \over d}\,,2
\left|\ln{\Gamma_0 \over \omega}\right|\right])\right)$$

\begin{equation}\left.-{3 \over 4}\sigma\cos(\omega t)\left(-d^{-1}(2\sigma)^{{\theta \over
d}-1}({\rm G}\left[1-{\theta \over d}\,,2 \left|\ln
(t\Gamma_0)\right|\right]-{\rm G}\left[1-{\theta \over d}\,,2
\left|\ln{\Gamma_0 \over
\omega}\right|\right])\right)\right)\end{equation} where ${\rm
G}[\alpha,x]$ is the incomplete gamma function.

If we use the asymptotic representation for the incomplete gamma
function for large values of the its second argument we obtain for
the difference $\Delta {\rm G}$ of two incomplete gamma functions

\begin{equation}\Delta {\rm G}\sim\left[1-{1 \over (\omega t)^2}
\left(\frac{|\ln(\omega^{-1} \Gamma_0)|}{\ln(t
\Gamma_0)}\right)^{1+{\theta \over d}}\right] {\omega^2 \over
\Gamma_0^2}.
\end{equation}

We observe a some similarity with expression (5.7) (in brackets)
in [35] in conformity with our quantum regime.

At derivation of (26) we made the following approximations: 1)
${\rm si}(b_L t)\simeq-(p_L t)^{-1}$ $\times \cos(p_L t)$; 2)
${\rm ci}(b_L t)\simeq(p_L t)^{-1} \sin(p_L t)$; 3) $b_L\simeq2
\omega+\Gamma_L$.

We also made the numerical calculation of the average over $L$ of
the expression (25) without these approximations and obtained
almost the same curves for susceptibility as ones obtained from
(26) (the parameter values we give below).

The susceptibility $\chi'(\omega,t)$ depends on many parameters of
the droplet system and the external $ac$ magnetic field, for
example, it depends on the kind of distribution function
$P_L(\epsilon_L)$, on the droplet microscopic tunneling rate
$\Gamma_0$, on time $t$, on frequency and amplitude of $ac$
magnetic field, on the droplet features.

The expression (26) consists of the time-independent terms, the
terms which describe the simple oscillations with frequency
$\omega$ and the terms which depend on time $t$ and determine
nonstationary nonequilibrium dynamics of droplet system. So the
real part of the $ac$ susceptibility can be represented
approximately as a sum of the stationary part ($\chi'_{ST}$) and
nonstationary part ($\chi'_{NST}$):

\begin{equation}\chi'(\omega,t)\simeq\chi'_{ST}+\chi'_{NST}.\end{equation}

Let us take for numerical calculation of expression (26) the
following values of parameters: $d=3$, $\theta=0.5$,
$\gamma=10^{-15}$, $\Gamma_0=10^8,\,10^{10},\,10^{12}$, $h=1$,
$\sigma=10^{-15}$, $t=0\div100$, $\omega=0.05,\,0.1$.

In Fig. 1 we show the $t$ - dependence of the real part of $ac$
susceptibility of the droplet system $\chi'_{NST}$. Here $\omega
t$ is comparable or more than unity, so one may observe
nonstationary dynamics and the aging regime [12]. In Fig. 1a it is
observed the slow dynamics at $\Gamma_0=10^8$, $\omega=0.05$,
$0.1$ and $t=0\div25$. At first the curve grows (very quickly) up
to some finite value and then falls down. At more long times the
$ac$ susceptibility shows a decay in queue. In Fig. 1b and Fig. 1c
the $t$ - dependence of $\chi'_{NST}(\omega,t)$ is shown for more
long times: b) $t=0\div50$; c)~$t=0\div100$. The time interval
covers two decades of the elapsed time $t$. We see that with
frequency increasing the susceptibility magnitude decreases and
the slow dynamics is suppressed at more high frequency. The $ac$
susceptibility tends more rapidly to zero at more long times. In
Figures 1 we see the essential influence of frequency $\omega$ and
elapsed time $t$ on the susceptibility.

In Fig. 2 we give the time dependence of $\chi'_{NST}(\omega,t)$
at different values of fixed quantum parameter $\Gamma_0$ (the
microscopic tunneling rate $\Gamma_0=10^8,\,10^{10},\,10^{12}$).
We observe the small effect of quantum parameter $\Gamma_0$ on
susceptibility $\chi'_{NST}(\omega,t)$. With increasing of
$\Gamma_0$ the magnitude of the susceptibility slightly decreases
at small times, i. e. the quantum fluctuations in some sense
decrease the susceptibility in quantum regime in spin glass phase.

We may compare our data with experimental ones [18, 24, 25] only
very approximately because in the $\chi'(\omega,t)$ - experiments
and in our paper the qualitatively different spin-glass systems
are considered. We observe qualitatively similar dynamical
behavior of $\chi'(\omega,t)$ in the range of the small times
elapsed since the quench.

\section{Discussion and conclusion}
In this paper we have investigated the low temperature
nonequilibrium dynamical behavior of the magnetic $ac$
susceptibility in $d$ - dimensional Ising spin glass with
short-range interactions between spins in a transverse field in
terms of the phenomenological droplet model. The real part of the
low-frequency $ac$ susceptibility $\chi'(\omega,t)$ as a function
of time $t$ (which the system waits from quench time moment till
the time moment of measurement) and the frequency $\omega$ of
external $ac$ magnetic field is calculated. We display the
nonequilibrium dynamics for different values of low frequency of
$ac$ field at constant temperature in spin glass phase. We can see
the oscillations of $\chi'(\omega,t)$ from expression (26). The
real part of $ac$ magnetic susceptibility $\chi'(\omega,t)$ of
droplet system at very low temperatures (quantum regime) has
two-time regions in which the time evolution is of a different
nature. At short times $t$ we observe nonequilibrium dynamical
behavior - the time decay of $\chi'(\omega,t)$ at low frequency
and at constant temperature $T$ ($\Gamma_L\gg k_B T$ in quantum
regime). If the frequency of $ac$ field increases the
nonequilibrium dynamics is suppresed. So, the response of droplet
system to external perturbating field depends on the thermal
history.

In [44] it was shown that the behavior of the response function
$R(t,t_w)$ demonstrates the existence of the stationary and aging
regimes in quantum systems. The theoretical curve (Fig. 2 in [44])
$R(t,t_w)$ was given as function of $\tau$ ($\tau=t-t_w$) if
$\tau\in[0,50]$ and $t_w=2.5,\,5,\,10,\,20$ and 40. For
$\tau\leq\tau_{char}$ ($\tau_{char}$ is a some characteristic
time) it was found the stationary regime, if $\tau>\tau_{char}$
the dynamics is nonstationary. In [18, 24, 25] the
$\chi'(\omega,t)$ - experimental data are represented. In aging
regime the slow monotonous decay of $\chi'(\omega,t)$ was
observed. In our quantum system at very low $T$ we can not find
good agreement with these data.

Besides we have shown that the quantum fluctuations have slight
influence to the dynamical susceptibility of droplet system at
very low temperatures. Their increase leads to the small decrease
of the susceptibility magnitude. Like in [43] it is shown that in
the aging regime of quantum spin glasses of rotors all terms in
the dynamical equations governing the time evolution of the spin
response and correlation function that arise solely from quantum
effects are irrelevant at long times. The quantum effects enter
only through the renormalization of the dynamical equations
parameters [43]. In paper [44] it is shown that quantum
fluctuations in quantum glassy systems depress the phase
transition temperature, in a glassy phase aging survives the
quantum fluctuations and the quantum fluctuation - dissipation
theorem is modified due to quantum fluctuations.

Thus, we have found the nonequilibrium low-frequency dynamical
behavior of the $ac$ susceptibility at constant temperature for
droplet system in quantum regime ($\Gamma_L\gg k_B T$) in spin
glass phase. For small times $t$ the magnetic $ac$ susceptibility
depends on both frequency of $ac$ magnetic field $\omega$ and on
time $t$ elapsed since the sample reached the temperature $T_1$
($T_1<T_g$). We do not find slow continues decrease of the
amplitude of $\chi'(\omega,t)$ as a function of the time $t$ at
long times. Aging is more visible in the out-of-phase component
$\chi''(\omega,t)$ of the magnetic susceptibility which determines
a dissipation [12]. A qualitative agreement with
$\chi'(\omega,t)$-measurements [18, 24, 25] in spin-glass systems
was found for small times and low frequencies.

This work is partially supported by the RBRF under Grant
01-02-16368.

\newpage
\centerline{\Large References}

\begin{enumerate}
\item P. G. de Gennes, Rev. Mod. Phys. 71, 5374 (1999)

\item E. R. Nowak, J. B. Knight, E. Ben-Naim, H. M. Jaeger and S.
R. Nagel, Phys. Rev. E 57, 1971(1998)

\item M. Nicokas, P. Duru and O. Pouliquen, Eur. Phys. J, E 3, 309
(2000)

\item H. M. Jaeger, C.-H. Liu and S. R. Nagel, Phys. Rev. Lett.
62, 40 (1989)

\item E. Falcon, R. Wunenburger, P. Evereque, S. Fauve, C. Chabot,
V. Garrabos and D. Beysens, Phys. Rev. Lett, 83, 440 (1999)

\item A. P. Young (Ed.) "Spin-glasses and random fields" (World Scientific, Singapore, 1998)

\item L. C. E. Struik "Physical aging in amorphous polimers and
other materials" (Elsevier. Houston. 1978)

\item D. Fisher and Pierre Le Doussal, Phys. Rev. E 64, 066107
(2001)

\item J.A. Mydosh "Spin Glasses: an experimental introduction"
(Taylor \& Francis, London, 1993)

\item R. L. Leheny and S. R. Nagel, Phys. Rev. B 57, 5154 (1998)

\item F. Alberci-Kios, J.-Ph. Bouchaud, L. F. Cugliandolo, P.
Doussineau and A. Levelut, Phys. Rev. Lett, 81, 4987 (1998)

\item E. Vincent, J. Hammann, M. Ocio, J.-Ph. Bouchaud and L. F.
Cugliandoloin in "Complex behavior of glassy systems". Ed.: M.
Rubi (Springer, Verlag, Berlin, 1997)

\item J.-Ph. Bouchaud, L. F. Cugliandolo, J. Kurchan and M. Mezard
in "Spin Glasses and Random Fields". Ed.: A. P. Young (World
Scientific, Singapore, 1998)

\item P. Nordblad and P. Svedlindh in "Spin Glasses and Random Fields". Ed.: A. P. Young (World
Scientific, Singapore, 1998)

\item D. A. Huse, Phys. Rev. B 43, 8673 (1991)

\item L. Bertier and J.-Ph. Bouchaud, cond-mat/0202069

\item P. E. Jonsson, H. Yoshima, P. Nordblad, H. Agura Katori and
A. Ito, preprint cond-mat/0112389

\item V. Dupuis, E. Vincent, J.-Ph. Bouchaud, J. Hammann, A. Ito
and H. Agura Katori, Phys. Rev. B 64, 174204(2001)

\item Y. G. Joh, R. Orbach, G. G. Wood, J. Hammann and E. Vincent,
preprint cond-mat/0002040

\item E. Vincent, V. Dupuis, M. Alba, J. Hammann and J.-Ph.
Bouchaud, Europhys. Lett. 50, 674 (2000)

\item V. S. Zoter, G. F. Rodriguez, R. Orbach, E. Vincent and J.
Hammann, preprint cond-mat/0202269

\item C. Djurberg, J. Mattsson and P. Nordblad, Europhys. Lett.
29, 163 (1995)

\item L. Lundgren, P. Nordblad, P. Svedlindh and O. Beckman, J.
Appl. Phys. 57, 3371 (1985)

\item A. G. Shins, A. F. Arts and H. W. de Wijn, Phys. Rev. Lett.
70,2340 (1993)

\item P. Svedlindh, K. Gunnarsson and J.-O. Andersson, Phys. Rev.
B 46, 13867 (1992)

\item P. Shukla and S. Singh, Phys. Rev. B 23, 4661 (1981)

\item T. K. Kopec, Phys. Rev. B 54, 3367, (1996)

\item T. Giamarchi and P. L. Doussal, Phys. Rev. B 53, 15206
(1996)

\item T. Nieuwenhuizen, Phys. Rev. Lett. 74, 4289 (1995)

\item J. Miller and D. A. Huse, Phys. Rev. Lett 70, 3147 (1993)

\item D. Grempel and M. Rozenberg, Phys. Rev. Lett. 80, 389 (1998)

\item H. Rieger and A. P. Young "Quantum spin-glasses"
(Springer-Verlag, Berlin, 1996)

\item J. Schwinger, J. Math. Phys. 2, 407 (1961);

L. V. Keldysh, Zh. Eksp. Teor. Fiz. 47, 1515 (1964)

\item C. Chamon, A. W. W. Ludwig and C. Nayak, Phys. Rev. B 60,
2239 (1999);

A. Kamenev and A. Andreev, Phys. Rev. B 60, 2218 (1999);

M. N. Kiselev and R. Oppermann, Phys. Rev. Lett. 85, 5631 (2000)

\item D. S. Fisher and D. A. Huse, Phys. Rev. Lett. 56, 1601 (1986); Phys. Rev. B
36, 8937 (1987); Phys. Rev. B 38, 373, 386 (1988)

\item M.J. Thill and D.A. Huse,  Physica A 241 , 321 (1995)

\item G. Busiello and R. V. Saburova, Int. J. Mod.  Phys. B 14, 1843 (2000)

\item G. Busiello, R. V. Saburova and V. G. Sushkova, Solid. St.
Commun. 119, 545 (2001); J. Phys. Studies, 5, 1 (2001)

\item T. Komori, H. Yoshino and H. Takayama, J. Phys. Soc. Jap.
68, 3387 (1999); 69, 1192 (2000)

\item H. Yoshino, K. Hukushima and H. Takayama, preprint
cond-mat/0202110

\item H. E. Castillo, C. Chamon, L. F. Cugliandolo, M. P. Kennett,
preprint cond-mat/0112272

\item V. S. Zotev and R. Orbach, preprint cond-mat/0112489

\item  M. P. Kennet, C. Chamon and J. Ye, Phys. Rev. B 64, 224408
(2001)

\item L. F. Cugliandolo and G. Lozano, Phys. Rev. Lett. 80, 4979
(1998); Phys. Rev. B 59, 915 (1999);

L. F. Cugliandolo, preprint cond-mat/0107596;

 H. E. Castillo, C. Chamon, L. F. Cugliandolo and M. P. Kennet, preprint
cond-mat/0112272

\item J. Kisker, L. Santen and M. Schreckenberg, Phys. Rev. B 53,
64181 (1996)

\item F. Krzakava and O. C. Martin, Phys. Rev. Lett. 85, 3013
(2000)

\item J. Houdayer and O. C. Martin, Europhys. Lett. 49, 794 (2000)

\item W. T. Grandy Jr. "Foundation of statistical mechanics" (D.
 Reidel Publishing Company, Dordrecht, Holland, 1988)

 H. J. Kreuzer "Nonequilibrium thermodynamics and its statistical
 foundations" (Charendon Press, Oxford, 1981)

\end{enumerate}

\newpage

\centerline{\Large Figure captions}

\bigskip

Fig. 1. The in-phase susceptibility $\chi'(\omega, t)$
(nonstationary part) as a function of time $t$ for quantum
parameter $\Gamma_0=10^8$ and fixed values of frequency
$\omega=0.05,\,0.1$:

a) $\omega=0.05,\,0.1;\,t=0\div25;$

b) $\omega=0.05,\,0.1;\,t=0\div50;$

c) $\omega=0.05,\,0.1;\,t=0\div100;$

\bigskip

Fig. 2. The in-phase susceptibility $\chi'(\omega, t)$
(nonstationary part) as a function of time $t$ for frequency
$\omega=0.1$ and for three values of quantum parameter
$\Gamma_0=10^8,\,\Gamma_0=10^{10},\,\Gamma_0=10^{12}$.

\end{document}